\documentclass[9pt,twocolumn,twoside]{opticajnl}
\journal{opticajournal} % use for journal or Optica Open submissions

\setboolean{shortarticle}{true}
\title{An ultra-stable cryogenic sapphire cavity laser with an instability of $1.9\times10^{-16}$ based on a low vibration level cryostat}

\author[*]{Leilei He, Jingxuan Zhang, Zhiyuan Wang, Jialu Chang, Qiyue Wu, Zehuang Lu, Jie Zhang}

\affil{MOE Key Laboratory of Fundamental Physical Quantities Measurement, \& Hubei Key Laboratory of Gravitation and Quantum Physics, PGMF and School of Physics, Huazhong University of Science and Technology, Wuhan 430074, P. R. China\\}

\affil[*]{jie.zhang@mail.hust.edu.cn}

\begin{abstract}
Cryogenic ultra-stable lasers have extremely low thermal noise limits and frequency drifts, but they are more seriously affected by vibration noise from cryostats. Main material candidates for cryogenic ultra-stable cavities include silicon and sapphire. Although sapphire has many excellent properties at low temperature, the development of sapphire-based cavities is less advanced than that of silicon-based. Using a homemade cryogenic sapphire cavity, we develop an ultra-stable laser source with a frequency instability of $1.9\times10^{-16}$. This is the best frequency instability level among similar systems using cryogenic sapphire cavities reported so far. Low vibration performance of the cryostat is demonstrated with a two-stage vibration isolation, and the vibration suppression is further improved by different mixing ratio of the gas-liquid helium. With this technique, vibrations at frequencies higher than tens of hertz are greatly suppressed.
\end{abstract}

\setboolean{displaycopyright}{false} % Do not include copyright or licensing information in submission.

\begin{document}

\maketitle
Ultra-stable lasers have been widely used in optical clocks \cite{brewer2019al+,oelker2019demonstration}, tests of relativity \cite{zhang2021127666}, and gravitational wave detection \cite{bailes2021gravitational}. The frequency instabilities of ultra-stable lasers are limited by the thermal noise limit of the optical reference cavities \cite{numata2004thermal}. A straight forward way to lower the thermal noise limit is to operate optical cavities in cryogenic environment. The most commonly used materials for cryogenic optical cavities include monocrystalline silicon and sapphire \cite{matei20171,robinson2019crystalline,talla2022laser,kudeyarov2021comparison,ushiba2015laser,wiens2020simplified,seel1997cryogenic,fluhr2016characterization}. Benefiting from the extremely low coefficient of thermal expansion (CTE) and low aging effect of these materials under cryogenic temperature, the cryogenic ultra-stable lasers also show extremely low frequency drifts \cite{robinson2019crystalline,wiens2020simplified,storz1998ultrahigh}.

Mono-crystalline silicon has multiple zero-crossing temperature points of CTE at cryogenic temperature, which are near 3.5 K, 17 K, and 124 K, respectively \cite{wiens2014silicon,wiens2020simplified}. The influence of external temperature fluctuations can be greatly reduced by setting the cavity temperature at these zero-crossing points. Cryogenic lasers stabilized to silicon cavities under 124 K and 4 K have demonstrated record frequency instabilities \cite{matei20171,robinson2019crystalline}. However, the narrow optical transparency range of silicon from 1.1 $\mu$m to 8 $\mu$m limits the direct applications of cryogenic lasers in diverse fields. For example, the most important application of ultra-stable lasers is in optical clocks, but almost all the operational optical clocks have their clock transition frequencies from ultraviolet to visible range, such as 267 nm for Al$^+$, 792 nm for Ca$^+$, and 698 nm for Sr, etc \cite{brewer2019al+,oelker2019demonstration,sanner2019optical, jian2022improved,huang2022liquid,rosenband2008frequency,ohtsubo2020frequency,  mcgrew2019towards, ohmae2020direct, wilpers2007absolute}. In addition, the commonly used 1064 nm light source for gravitational wave detection is also out of the silicon transmission window. Although the laser frequency can be indirectly stabilized using an optical frequency comb as a transfer bridge, the whole system is more expensive and complex. 

By contrast, the optical transparency range of sapphire is from 0.15 $\mu$m to 5.5 $\mu$m, which conveniently meets the requirements of most optical applications. Moreover, sapphire crystal has a higher Young's modulus (464 GPa) than silicon (188 GPa), which makes the sapphire cavity naturally less sensitive to vibrations. Furthermore, sapphire crystal has a high thermal conductivity (280 W/m/K), which ensures better temperature uniformity of the cavity \cite{wolfmeyer1971thermal}. Compared with amorphous material like ultra-low expansion (ULE) glass, the frequency drift caused by stress relaxation of sapphire crystal is negligible \cite{storz1998ultrahigh}.

In this letter, we report on a 1070 nm ultra-stable laser locked to a cryogenic sapphire cavity at 3.7 K, with its fourth harmonic serving as the clock laser of an Al$^+$ optical frequency standard. We improve system performance by considering major technical noise sources. A high finesse sapphire cavity is developed with a large phase discriminator slope to reduce the noise of locking system. The long-term drift of the cavity length is reduced using specially designed thermal shields and active optical power stabilization. To suppress vibration noise, the main noise in cryogenic systems, we control the mixing ratio of gas-liquid helium in a custom-designed closed-cycle cryostat for optimum vibration isolation. With these improvements, the achieved fractional frequency instability of the laser is $1.9\times10^{-16}$ at 20 s, which is an order of magnitude better than the previous best result based on a cryogenic sapphire cavity \cite{seel1997cryogenic}.

\begin{figure}
\centering\includegraphics[width=0.4\textwidth]{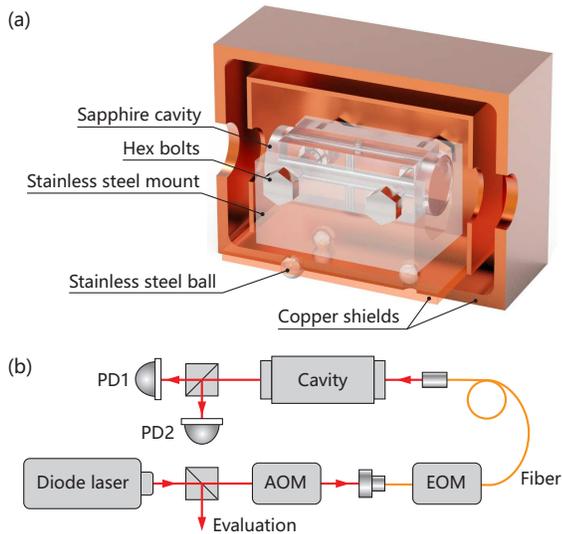}
\caption{\label{fig:1}(a) The mounting structure of the sapphire cavity. (b) The experimental setup of the cryogenic ultra-stable laser system. AOM, acousto-optic modulator; EOM, electro-optical modulator; PD, photodiode.}
\end{figure}

The experimental setup of the cryogenic ultra-stable laser system is shown in Fig. \ref{fig:1}. A 6 cm long homemade all-sapphire cavity sits inside a closed-cycled cryostat. The cavity is surrounded with active/passive temperature control shields. An 1070 nm diode laser is locked to the cavity using the Pound-Drever-Hall (PDH) method. As the core element in the cryogenic ultra-stable laser system, a high finesse cavity is essential for laser frequency stabilization. Sapphire is very hard, and ranks 9 on Mohs hardness scale with 10 the hardest for diamond. Therefore, it is difficult to polish sapphire to reach the specified surface flatness and smoothness. Special polishing techniques are adopted to ensure the surface flatness is better than $\lambda/10$ and parallelism better than $0.5^{\prime \prime}$. By selecting the best available high reflectivity cavity mirror pair, the finesse of the homemade sapphire cavity is measured to be about $4.9\times10^{5}$ using the cavity ring down method. It shall be noted that if the loss of the front and rear cavity mirrors are not matched, the frequency locking noise of the cavity stabilized lasers increases due to impedance mismatch. Therefore, we select the cavity mirror pair with an appropriate loss ratio. With an average 0.999997 reflectivity of the mirrors and a 2 ppm front cavity mirror loss, the impedance matching coefficient of our homemade cavity is 0.33, which agrees well with the theoretical value \cite{shi2018suppression}. With this cavity, we reach a 78\% coupling efficiency, which affects the phase discriminator slope and the frequency locking noise. By modulating the input error signal of the PDH servo system, the phase discriminator slope is measured to be 64 $\mu$V/Hz with a 30 $\mu$W incident optical power, showing the promise of a mHz level locking \cite{ye2021vibration,zhang2016characterization}.

Although the CTE of the sapphire crystal does not have a zero-crossing point, it is proportional to the power cube of the temperature, which is rather small under cryogenic temperature \cite{taylor1996measurement,ye2021investigation}. Operating the sapphire cavity at very low temperature can reduce the cavity length sensitivity to fluctuations of the environmental temperature and the incident laser power. We control the temperature at 3.3 K, which is the lowest temperature that the cryostat can reach to reduce the temperature fluctuation effect. The CTE of the sapphire crystal is $2.7\times10^{-11}$/K at 3.3 K. To achieve a target frequency instability at the level of $10^{-17}$, the temperature fluctuation of the cavity should be less than $10^{-6}$ K. The resolution of the best commercial available temperature sensor is 0.1 mK at sub-10 K, which limits the temperature stability of the active control. A thermal time constant over thousands of seconds between the active temperature controlled shield and the cavity is essential to ensure that the temperature fluctuation of the cavity can meet the requirement. The cavity sits on top of a stainless steel block, and four screws are inserted into the cavity to hold the cavity, as shown in Fig. \ref{fig:1}(a). To reduce the heat conductivity from the support structure, three stainless steel balls are used as heat breaks at the bottom of the copper block. The measured thermal time constant is 9550 s.

\begin{figure}
\centering\includegraphics[width=0.4\textwidth]{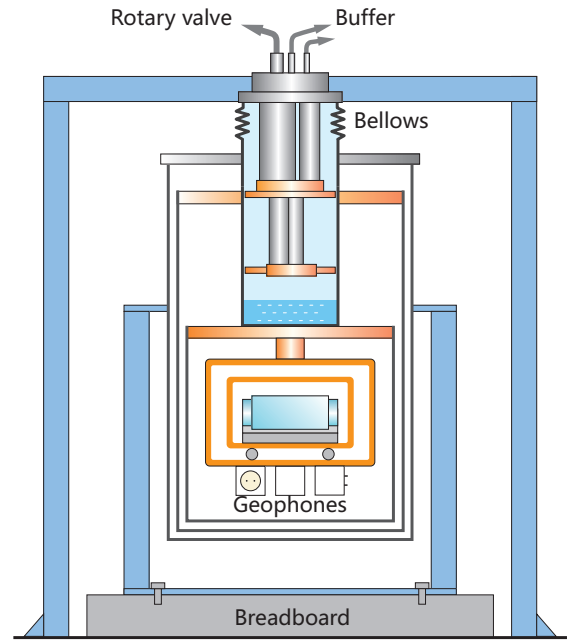}
\caption{\label{fig:2}Schematic of the cryocooler system.}
\end{figure}

Sapphire crystal has a small heat capacity at cryogenic temperature. In addition to the external temperature fluctuation, the incident laser power fluctuation can also cause temperature variation of the cryogenic sapphire cavity. As shown in Fig. \ref{fig:1}(b), we actively control the output laser power from the cavity use the PD1 as the power sensor and compensate the power jitters using the AOM. The level of power jitters after active servo control is monitored by the outer-loop sensor PD2, the contribution of the laser power jitters after servo control is less than the cavity thermal noise limit.

In order to continuously operate the laser system, we choose a pulse-tube (PT) cryocooler as the closed-cycle cryostat. Although the PT cryocooler has no moving parts in the cold head, the vibration noise induced by the cryocooler is still one of the main noise sources, limiting the development of cryogenic ultra-stable laser systems. Passive or active techniques have been reported by several groups for further vibration suppression, such as keeping the vibration source away from the experimental area, using soft connection springs, gas buffer, or a liquid/gas mixing zone between the cold head and the working area, and applying active vibration control, etc \cite{d2018active,wang2010vibration,millo2014cryogenic,wiens2021simple,caparrelli2006vibration,zhang2017ultrastable,ushiba2015laser,dubielzig2021ultra,micke2019closed}.

We pay great care to design the cryostat in order to reach a low vibration level \cite{ye2021vibration}. The schematic of the designed cryocooler system is shown in Fig. \ref{fig:2}. 
We place the laser system inside a cave lab with extremely low temperature fluctuations and vibration level. The compressor of the cryocooler is housed in a separate soundproof room to isolate vibration and acoustic noise. The compressor and the rotary valve are connected by two 20 m long metal tubes, which are rigidly fixed to the wall to reduce vibration effect. A suspended cryostat, mounted on a bulky and rigid shelf, is used to decrease the dimension of the apparatus and avoid heat loss of the cooling rod. Damped by metal bellows, the bottom of the cryostat chamber is fixed to the cave ground. The sapphire cavity is placed on the hanging structure under the second cooling stage.

\begin{figure}
\centering\includegraphics[width=0.4\textwidth]{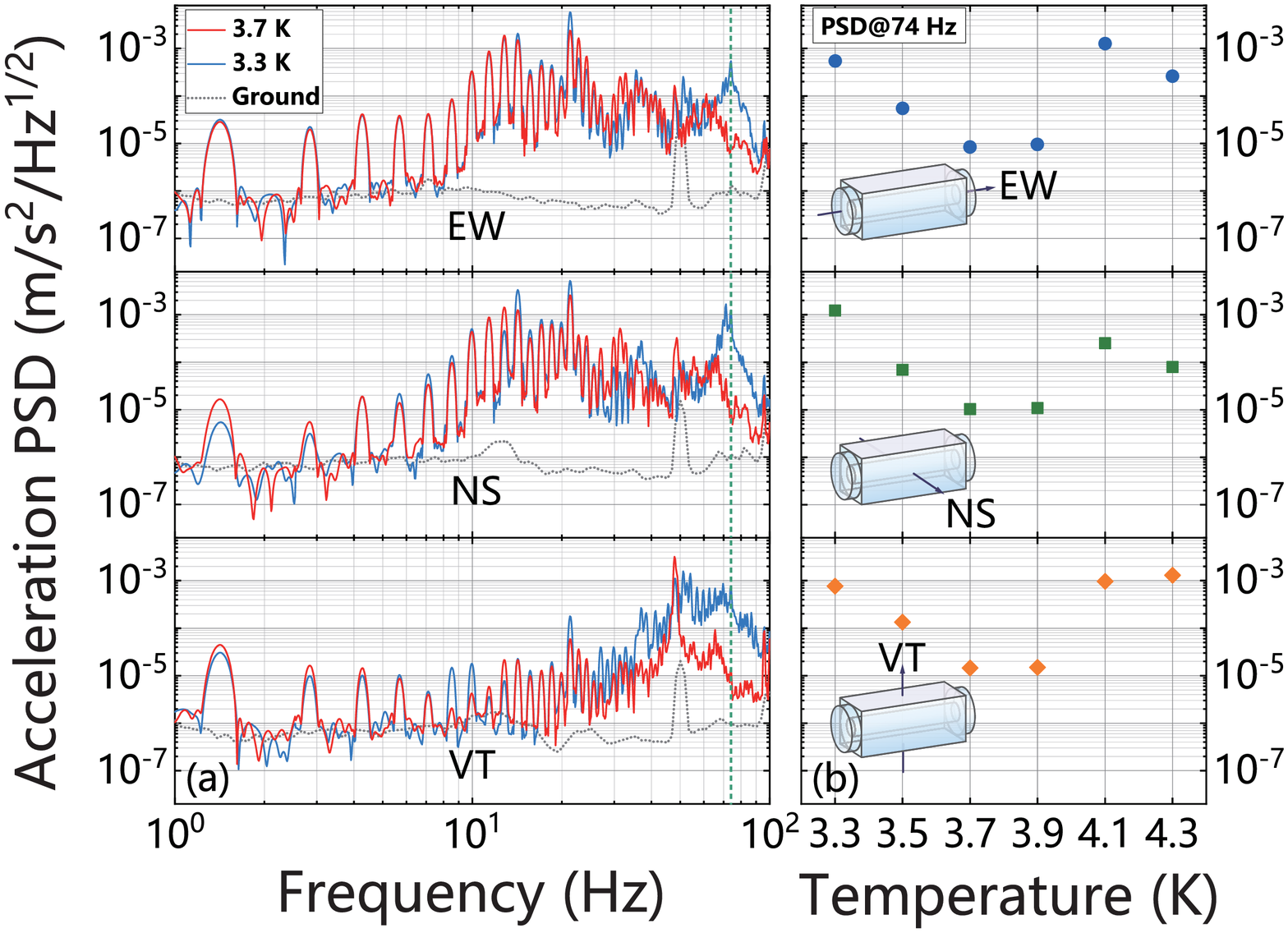}
\caption{\label{fig:3}(a) The measured vibration levels at the cavity housing area in directions of EW, NS and VT at 3.7 K (red line) and 3.3 K (blue line), compared with the vibration level on the cave ground (gray shadow). (b)  Vibration levels at the frequency of 74 Hz in three directions from 3.3 K to 4.3 K with different ratios of gas-liquid helium mixing.}
\end{figure}  

The main vibration source of the system is the rotary valve, which is the only moving component. The helium pressure waves are another vibration source, which lead to a periodic elastic deformation of the two coupled thin-walled cooling tubes. Pressure waves excite structural resonance of the system and worsen the performance of the cavity-stabilized laser. We suppress the vibration using liquid/gas helium mixture and measure with three in-situ calibrated vibration sensors in our system. The temperature is controlled below the boiling point of helium, and a thin layer of liquid helium is formed at the bottom of the gas chamber to reduce the deformation on the metal disk caused by the impact of the pressure waves. The layer thickness of the liquid helium and the pressure in gas chamber determine the vibration attenuation ability. Under a constant volume, the mixing ratio of the gas-liquid helium changes at different temperatures. We vary the temperature of the cold plate, and demonstrate an optimization of vibration level suppression under different operation parameters.

The measured vibration spectrum is shown in Fig. \ref{fig:3}. Fig. \ref{fig:3}(a) shows the vibration levels at the cavity housing area in directions of east-west (EW), north-south (NS) and vertical (VT), and the directions are defined relative to the sapphire cavity. The temperature is controlled from 3.3 K to 4.3 K with a step size of 0.2 K. The lowest temperature of the cryostat is 3.3 K, and at 4.3 K the confined liquid/gas helium mixture space is filled with helium gas. It is obvious that the vibration levels change at different temperatures, and an optimal vibration level in three directions can be achieved at a suitable temperature. Comparing with different ratios of gas-liquid helium mixture, vibration levels are significantly higher in VT direction at 4.3 K. In fact, the high frequency vibrations above tens of hertz are effectively suppressed at 3.7 K. We show two orders of magnitude vibration suppression at a specific frequency of 74 Hz in three directions at different temperature in Fig. \ref{fig:3}(b). Here 74 Hz is one of the mechanical resonant frequency of the cryostat, and can magnify the vibrational effect on the laser instability. The vibration levels also touch the levels of ground vibration (gray line) in the frequency range from 1 Hz to 10 Hz except the 1.4 Hz and its harmonic caused by the PT rotary valve. Vibrations are suppressed to the order of $10^{-7}$ m/s$^{2}$/Hz$^{1/2}$ at all three directions. This result is superior to most of the reported vibration levels of the PT cryocoolers \cite{wiens2021simple,zhang2017ultrastable,chijioke2010vibration,tomaru2004vibration,dubielzig2021ultra,caparrelli2006vibration,micke2019closed,ushiba2015laser}.

\begin{figure}
\centering\includegraphics[width=0.4\textwidth]{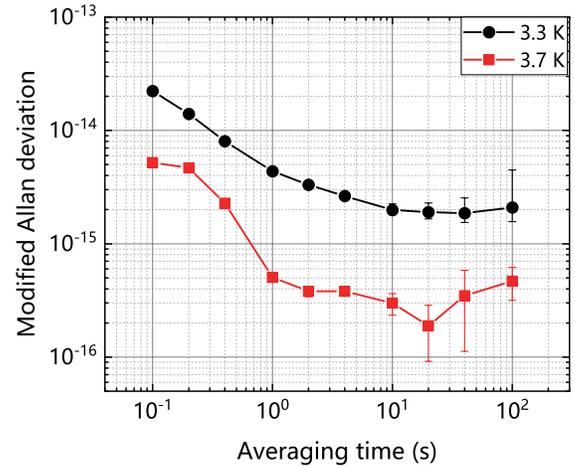}
\caption{\label{fig:4}The frequency instabilities of the laser based on the cryogenic sapphire cavity. Black curve is measured at 3.3 K, and red curve is measured at 3.7 K. It can be seen that the performance of the cyrogenic laser is greatly improved at 3.7 K due to better suppression of the vibration levels.}
\end{figure}

The frequency instability of the cryogenic laser is evaluated using the three-cornered-hat method. The two other reference lasers are locked to a 10-cm long ULE cavity and a 30-cm long ULE cavity, respectively, at room temperature with a frequency instability level of $10^{-16}$ \cite{zeng2018thermal,wang2021single,ma2020investigation}. The derived frequency instability of the cryogenic laser is shown in Fig. \ref{fig:4}, with the controlled temperature of the sapphire cavity at 3.3 K and 3.7 K, respectively. All other parameters are the same. At the optimum temperature of 3.7 K, the fractional frequency instability of the cryogenic laser reaches $1.9\times10^{-16}$ at 20 s averaging time, which is more than ten times better than the previous best result of cryogenic sapphire cavity. It can be seen that the performance of the cyrogenic laser is greatly improved at 3.7 K due to better suppression of the vibration levels.

\begin{backmatter}
\bmsection{Funding} This research was funded by the National Key R\&D Program of China (Grants No. 2022YFC2204002, and 2017YFA0304400), the National Natural Science Foundation of China (Grants No. 91336213, 61875065, 12004124, 11904112, and 11774108), and the Key-Area Research and Development Program of GuangDong Province (Grant No. 2019B030330001).

\bmsection{Disclosures} The authors declare no conflicts of interest.

\bmsection{Data Availability Statement} Data underlying the results presented in this paper are not publicly available at this time but may be obtained from the authors upon reasonable request.

\end{backmatter}

% Bibliography
\bibliography{sample}

% Full bibliography added automatically for Optics Letters submissions; the following line will simply be ignored if submitting to other journals.
% Note that this extra page will not count against page length
\bibliographyfullrefs{sample}

%Manual citation list
%\begin{thebibliography}{1}
%\bibitem{Zhang:14}
%Y.~Zhang, S.~Qiao, L.~Sun, Q.~W. Shi, W.~Huang, %L.~Li, and Z.~Yang,
 % \enquote{Photoinduced active terahertz metamaterials with nanostructured
  %vanadium dioxide film deposited by sol-gel method,} Opt. Express \textbf{22},
  %11070--11078 (2014).
%\end{thebibliography}

% Please include bios and photos of all authors for aop articles

\end{document}